\documentclass[12pt,pdf,aoas,preprint]{imsart}
\RequirePackage[OT1]{fontenc}
\RequirePackage{amsthm,amsmath}
\usepackage[authoryear]{natbib}
\RequirePackage[colorlinks,citecolor=blue,urlcolor=blue]{hyperref}
\RequirePackage[T1]{fontenc}
\RequirePackage{color}
\RequirePackage{graphicx}
\RequirePackage{comment}
\usepackage[margin=1.0in,letterpaper]{geometry}
\usepackage{numprint}
\usepackage{amssymb}
\usepackage{atbegshi,picture}

\startlocaldefs
\numberwithin{equation}{section}
\theoremstyle{plain}
\endlocaldefs
\newcommand*\cpp{C\kern-0.2ex\raisebox{0.4ex}{\scalebox{0.8}{+\kern-0.4ex+}}}
\newcommand{\slfrac}[2]{\left.#1\middle/#2\right.}

\DeclareMathOperator*{\med}{med}

\DeclareMathOperator*{\ave}{ave}

\expandafter\def\expandafter\normalsize\expandafter{%
    \normalsize
\setlength\abovecaptionskip{-.05cm}
}
\definecolor{fmcd}{RGB}{255,0,0}
\definecolor{fmve}{RGB}{0,0,255}
\definecolor{RCS}{RGB}{0,255,0}

\begin{document}

\begin{frontmatter}
\title{Finding Regression Outliers With FastRCS}
\runtitle{The RCS Outlyingness Index}
%\thankstext{T1}{Footnote to the title with the ``thankstext'' command.}

\begin{aug}
\author{\fnms{Kaveh} \snm{Vakili}\ead[label=e1]{kaveh.vakili@wis.kuleuven.be}} \and
\author{\fnms{Eric} \snm{Schmitt}\ead[label=e2]{eric.schmitt@wis.kuleuven.be}}
%\and
%\author{\fnms{Third} \snm{Author}\thanksref{t1,m2}
%\ead[label=e3]{third@somewhere.com}
\ead[label=u1,url]{http://wis.kuleuven.be/stat/robust}

%\thankstext{t1}{Some comment}
%\thankstext{t2}{First supporter of the project}
%\thankstext{t3}{Second supporter of the project}
\runauthor{K. Vakili and E. Schmitt}

\bigskip

%\affiliation{KU Leuven\thanksmark{m1}}

\address{%Statistics Section, Department of Mathematics\\
%Celestijnenlaan 200b - box 2400\\
%3001 Heverlee\\
\printead{e1}\\
\phantom{E-mail:\ }\printead*{e2}\\
%\printead{u1}
}

\begin{comment}%if mia.
\address{
%Statistics Section, Department of Mathematics\\
%Celestijnenlaan 200b - box 2400\\
%3001 Heverlee\\
\printead{e1}\\
\phantom{E-mail:\ }\printead*{e2}\\
%\printead{u1}
\end{comment}

%\address{Address of the Third author\\
%Usually a few lines long\\
%Usually a few lines long\\
%\printead{e3}\\

\end{aug}

\begin{abstract}
The Residual Congruent Subset (RCS) 
is a new method for finding outliers
 in the regression setting.
Like many other outlier detection 
procedures, RCS searches for a subset
 which minimizes a criterion. 
The difference is that the new criterion
 was designed to be insensitive to the 
outliers.
 RCS is supported by FastRCS, a fast regression
 and affine equivariant algorithm which we also 
detail. 
Both an extensive simulation study and two real
 data applications show that FastRCS performs 
better than its competitors.
\end{abstract}

%\begin{keyword}[class=AMS]
%\kwd[Primary ]{60K35}
%\kwd{60K35}
%\kwd[; secondary ]{60K35}
%\end{keyword}


\end{frontmatter}

\section{Introduction}
Outliers are observations that depart  
from the pattern of the majority of 
the data.
Identifying outliers is a major concern 
in data analysis for at least two reasons.
First, because a few outliers, if left unchecked, 
will exert a disproportionate pull on the fitted 
parameters of any statistical model, preventing the
analyst from uncovering the main structure in the data. 
Additionally, one may also want to find outliers 
to study them as objects of interest in 
their own right.
In any case, detecting outliers when there 
are more than two variables is difficult
 because we can not inspect the data visually and
 must rely on algorithms instead.

Formally, this paper concerns itself with the
 most basic variant of the  outlier 
detection problem in the regression context.
The general setting is that of the ordinary linear model:
\begin{equation}\label{mcs:lm}
y_i=\alpha+\pmb{x}_i'\pmb{\beta}+\epsilon_i\!
\end{equation}  
 where $\pmb{x}_i\in\mathbf{R}^{p-1}$ and
 $y_i\in\mathbf{R}$ have continuous distributions,
  $\epsilon_i\sim\mathrm{i.i.d.}\;\mathcal{N}(0,\sigma^2)$
 and $\pmb\theta:=(\alpha,\pmb\beta)$. Then, given a $p$-vector 
$\tilde{\pmb\theta}=(a,\pmb{b})$, we will denote the residual 
distance of $\pmb{x}_i$ to $y_i$ as:
\begin{equation}
r_i(\tilde{\pmb\theta})=|y_i-a-\pmb{x}'_i\pmb{b}|.
\end{equation}
We have a sample of $n$ observations $(\pmb x_i,y_i)$ 
with $n>p$, at least $h=\lceil(n+p+1)/2\rceil$ of 
which are well fitted by Model $\eqref{mcs:lm}$ and
our goal is to identify reliably the remaining ones. 
A more complete treatment of this
 topic can be found in textbooks \citep{mcs:MMY06}.

In this article we introduce
 RCS, a new procedure for 
finding regression outliers.
We also detail FastRCS, a 
fast  algorithm for computing it. 
The main output of FastRCS is an
 outlyingness index measuring how
 much each observation departs 
from the linear model fitting the majority
 of the data. 
The RCS outlyingness index is affine
 and regression equivariant (meaning 
that it is not affected by transformations
 of the data that do not change the ranking
 of the squared residuals) and can be computed 
efficiently for moderate values of 
$p$ and large values of $n$.
For easier outlier detection 
problems, we find that 
 FastRCS yields similar results 
as state of the art outlier 
detection algorithms. 
When considering more difficult
 cases however we find that the
 solution we propose leads to 
 much better outcomes. 

In the next section we
 motivate and define the RCS
 outlyingness and FastRCS. 
Then, in Section 3 we compare FastRCS 
to several competitors on synthetic data. 
In Section 4 we conduct two real data
 comparisons.

\section{The RCS outlyingness index}
\subsection{Motivation}
Given a sample of $n$ potentially contaminated 
observations $(\pmb x_i,y_i)$, 
the goal of FastRCS is to reveal the outliers. 
It is well known that this problem is also 
equivalent to that of finding a fit of Model $\eqref{mcs:lm}$ 
 close to the one we would have found without the outliers. 
Indeed, to ensures that they stand out in 
 a plot of the fitted residuals, 
it is necessary to prevent the outliers from 
pulling the fit in their direction.
Other equivariant algorithms that share the 
same objective are FastLTS \citep{mcs:RV06}
and FastS \citep{mcs:SY06}.

\begin{figure*}[h!]
\centering
\includegraphics[width=1\textwidth]{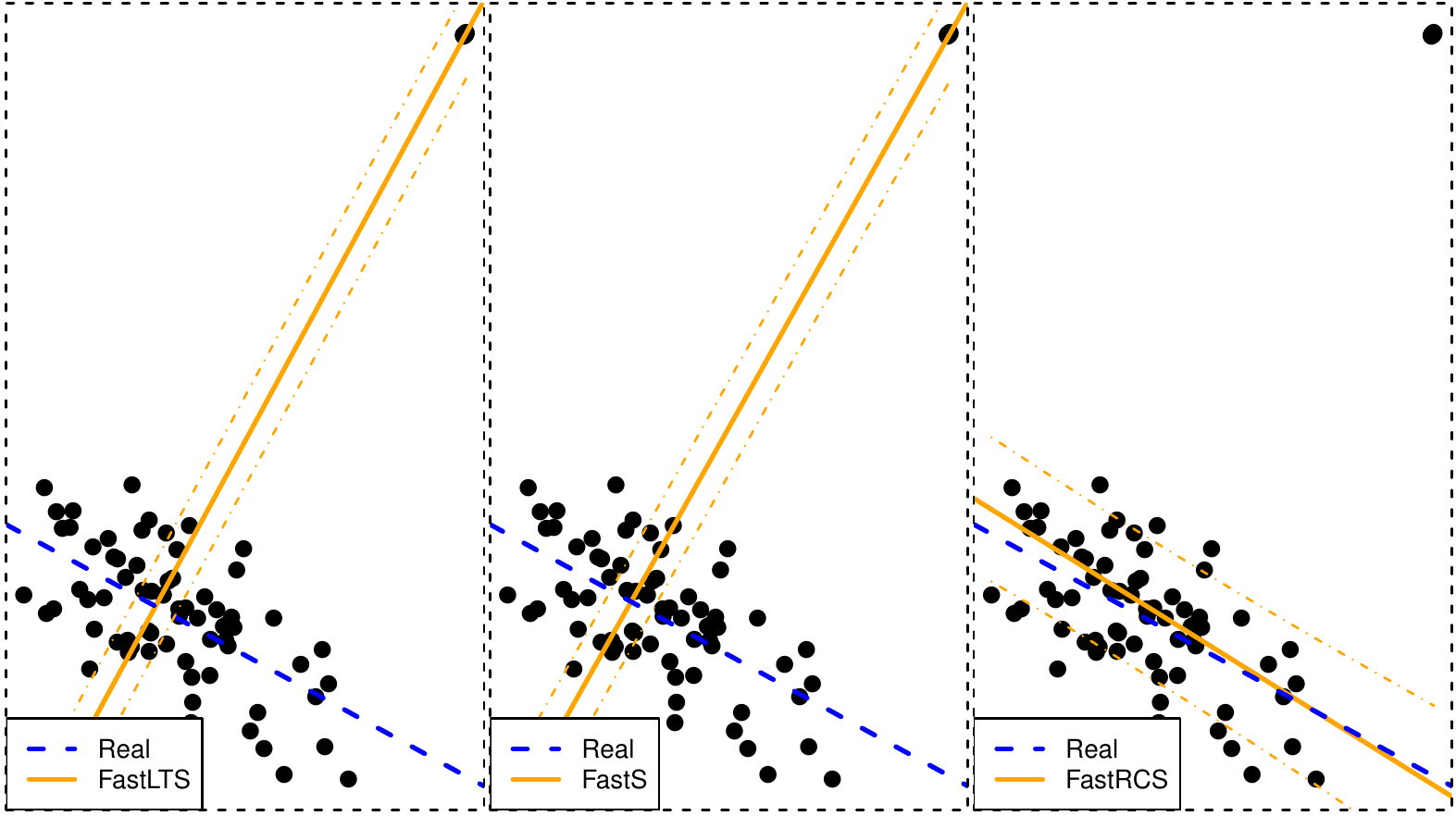}
	\caption{The three panels depict the 
same data-set. In each panel, the darker blue, dashed line 
shows the model governing the distribution of the 
majority --here 70 out of 100-- of the observations. The solid orange line
 shows, respectively, the vector $\theta^*$ fitted by each algorithm.} 	
\label{mcs:f1a}
\end{figure*}

However, in tests and real data examples, we 
often encounter situations where the outliers 
have completely swayed the 
fit found by FastLTS and FastS 
 yielding models that do not faithfully 
 describe the multivariate pattern of the bulk 
 of the data. Consider the following example.
 The three panels in Figure~\ref{mcs:f1a} depict
 the same 100 data points $(\pmb{x}_i,y_i)$: 70
 drawn from Model $\eqref{mcs:lm}$ 
and 30 drawn from a concentrated cluster of observations.
 The orange, solid lines in the first two panels 
 depict, respectively, the line corresponding to the  
 fit found by FastLTS (left) and FastS (center), 
both computed using the \texttt{R} package 
\texttt{robustbase} \citep{mcs:Ra12} with 
default parameters   
(the dashed orange lines depict the 95\% prediction
 intervals).
In both cases, the fits depicted in the first two
panels do not adequately describe 
--in the sense of Model $\eqref{mcs:lm}$-- 
 the pattern governing the distribution of the 
 majority of the data. This is because the outliers
 have pulled the fits found by FastLTS and FastS so
 much in their directions that their distances to 
 it no longer reveals them.

A salient feature of the algorithm we propose is
its use of a new measure we call the $I$-index 
(which we detail in the next section) to select
(among many random such subsets) an $h$-subset 
of uncontaminated data. The $I$-index 
characterizes the degree of spatial 
 cohesion of a cloud of points and its 
 main advantage lies in its
 insensitivity to the configuration of 
the outliers. As we
 argue below, this
makes the FastHCS fit as well
 as the the outlyingness index derived from it
 more reliable.

\subsection{Construction of the RCS Outlyingness index}

The basic outline of the algorithm is as follows. Given an $n$ by $p$ data matrix, FastRCS starts by drawing $M_p$ random subsets ('$(p+1)$-subsets'), 
denoted $\{H^m_0\}_{m=1}^{M_p}$,  
each of size $(p+1)$ of $\{1,\ldots,n\}$. 
Then, the algorithm grows each $H^m_0$ into a corresponding $H^m$, a subset of size $h$ of $\{1,\ldots,n\}$ (the letter $H$ without a subscript will always denote a subset of size $h$ 
of $\{1,\ldots,n\}$). The main innovation of our approach lies
in the use of the $I$-index, a new measure we detail below, 
to characterize each of these $H^m$'s. Next, FastHCS selects $H^*$, the 
 $H^m$ having smallest $I$-index. Then, the $h$ observations with indexes 
in $H^*$ determine the so-called raw FastRCS fit. Finally, we apply 
a one step re-weighting to this raw FastRCS fit to get the final FastRCS fit.
Our algorithm depends on two additional parameters ($K$ and $M_p$) but for 
clarity, these are not discussed in detail until Section~\ref{hcs:S24}. 

We begin by detailing the computation of the
 $I$-index for a given subset $H^m$. Denote 
$\tilde{\pmb\theta}^{mk}$ 
 the coefficients of the hyperplane 
(the index $k$ identifies these directions)
through $p$ data-points from $H^m$ (we detail below how 
we pick these $p$ data-points) and 
$H^{mk}$ the set of indexes of the $h$ data-points 
with smallest values of 
$r_i^2(\tilde{\pmb\theta}^{mk})$:
\begin{equation}
H^{mk}=\{i:r_{i}^2(\tilde{\pmb \theta}^{mk})\leqslant r_{(h)}^2(\tilde{\pmb \theta}^{mk})\}
\end{equation}
where $x_{(h)}$ denotes the $h$-th order statistic of a vector $\pmb x$. Then, we define the {\em incongruence index} of $H^m$ 
along $\tilde{\pmb\theta}^{mk}$ as:
\begin{equation}\label{mcs:crit1}
   I(H^m,\tilde{\pmb\theta}^{mk}):=
    \log\frac{\displaystyle\ave_{i\in H^m}r_i^2(\tilde{\pmb\theta}^{mk})
            }{\displaystyle\ave_{i\in H^{mk}}r_i^2(\tilde{\pmb\theta}^{mk})},
\end{equation}
with the convention that $\log(0/0):=0$. 
This index is always positive
 and will have small value if 
 the vector of $r_i^2(\tilde{\pmb\theta}^{mk})$ 
of the members of $H^m$  greatly 
overlaps with that of the members of
 $H^{mk}$. 
To remove the dependence of Equation
 \eqref{mcs:crit1} on  $\tilde{\pmb\theta}^{mk}$, we 
measure the incongruence of $H^m$ by considering
 the average over many directions:
\begin{equation}\label{mcs:crit2}
      I(H^m):=\ave_{\tilde{\pmb\theta}^{mk}\in B(H^m)} I(H^m,\tilde{\pmb\theta}^{mk})\;,
\end{equation}
where $B(H^m)$ is the set of all regression hyperplanes 
  through $p$ data-points with indexes in $H^m$. 
We call the $H^m$ with smallest $I(H^m)$ 
the {\em residual congruent subset} and 
 denote the index set of its members as 
$H^*$. Next, the raw FastRCS estimates are  
the parameters $(\pmb\theta^*,\sigma_*^2)$ 
fitted by OLS to the observations with 
indexes in $H^*$.

In essence, the $I$ index characterizes 
 the homogeneity of the members of a 
 given $h$-subset $H^m$ in terms of how
 much their residuals overlap with those of the
 $H^{mk}$ over many random regressions. %K% ordering as in PCS  
In practice, it would be  too laborious
 to evaluate Equation \eqref{mcs:crit2}
 over all members of $B(H^m)$.
A practical solution is to take
 the average over a random sample of $K$
 hyperplanes $\tilde{B}_K(H^m)$ instead.
The $I$-index is based on the observation
 that when the datum with indexes in $H^m$ 
form an homogeneous cloud of points,  
$\#\{H^m\cup H^{mk}\}$ tends to be large over many 
projection $a^{mk}$, causing $I(H^m)$ to be smaller. 

\begin{figure*}[h!]
\centering
\includegraphics[width=1\textwidth]{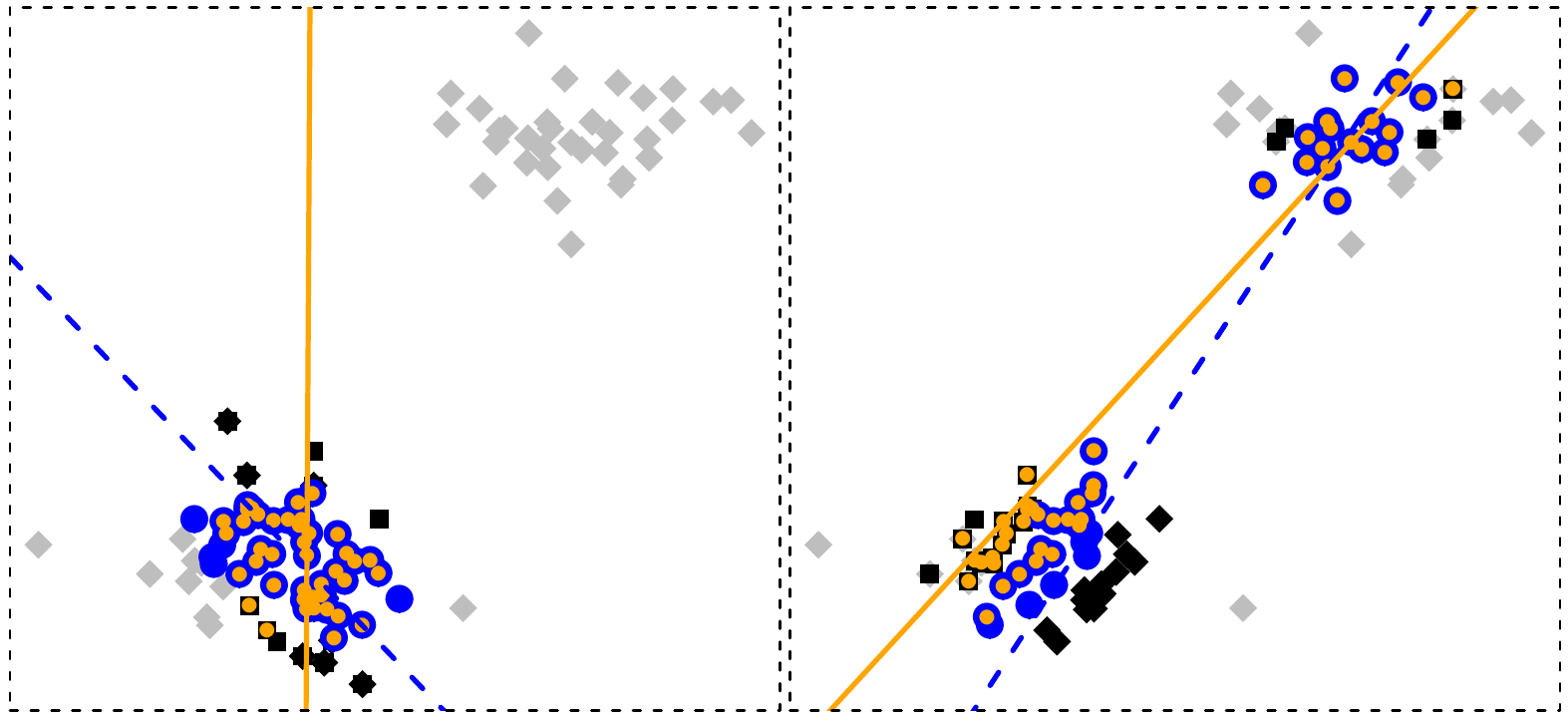}
	\caption{Incongruence index for a subset $H^1$ of homogeneous observations (left) and 
a subset $H^2$ of heterogeneous ones (right).} 	
\label{mcs:f1b}
\end{figure*}

 Consider the example shown in 
Figure~\ref{mcs:f1b}. Both panels depict
 the same set of $n=100$ data-points $(x_i,y_i)$. These 
points form two separate cluster. The main group contains 
70 points and is located on the left hand
 side. Each panel illustrates the behavior
 of the $I$ index for a given $h$-subset of
 observations. $H^1$ (left) forms a set of 
  homogeneous observations all drawn from the 
 same cluster. 
$H^2$, in contrast, is composed of data points
 drawn from the two disparate clusters. For each 
$H^m$-subset, $m=\{1,2\}$, we drew two 
regression lines $\tilde{\pmb\theta}^{m1}$ (dark blue, dashed) 
and $\tilde{\pmb\theta}^{m2}$ (light orange). 
The dark blue dots show the members of $\{H^m\cup H^{m1}\}$. 
Similarly, light orange dots show the 
members of $\{H^m\cup H^{m2}\}$.  The diamonds (black squares) show the members of $H^{m1}$
 ($H^{m2}$) that do not belong to $H^m$. After 
just two regression, the number of non-overlapping 
residuals (i.e.\{ $\{H^m\setminus H^{m1}\}\cap \{H^m\setminus H^{m2}\}\}$) 
 is 10 ($m=1$) and 21 ($m=2$) respectively. 
As we increase the number of regression lines   
$\tilde{\pmb\theta}^{mk}$, 
this pattern repeats and the difference between
 an $h$-subset that contains the indexes of an homogeneous
 cloud of points and one that does not grows steadily. 

For a given $h$-subset $H^m$, the $I$ index measures the typical 
size of the overlap between the members of $H^m$ and those 
of $H^{mk}$. 
Given two vector of coefficients 
$\tilde{\pmb\theta}^{m1}$, $\tilde{\pmb\theta}^{m2}$, 
the members of $H^{m1}$ and $H^{m2}$ not in $H^m$ (shown
 as diamonds and black squares in Figure~\ref{mcs:f1b})
 will decrease the denominator in Equation $\eqref{mcs:crit1}$
 without affecting the numerator, increasing the overall ratio. 
Consequently, $h$-subsets whose members 
 form an homogeneous cloud of points will have smaller 
values of the $I$ index. Crucially, the 
$I$ index characterizes an  $h$-subset composed 
 of observations forming an homogeneous cloud of 
 points independently of the 
configuration of the outliers. For example, the 
pattern shown in Figure~\ref{mcs:f1b} would still
 hold if the cluster of outliers were more 
 concentrated. 
This is also illustrated in the third
 sub panel in Figure~\ref{mcs:f1a} where the 
  parameters fitted by FastRCS are not unduly attracted
 by members of the cluster of concentrated outliers located 
on the right. 				%K%

In Sections \ref{mcs:s3} and \ref{mcs:s4}, we 
show that this new characterization allows FastRCS
 to reliably select uncontaminated $h$-subsets. 
This includes many situations where 
competing algorithms fail to do so. 
 First though, the following section details the 
FastRCS algorithm.

\subsection{A Fast Algorithm for the RCS Outlyingness}\label{hcs:S24}.

To compute the RCS outlyingness index, 
we propose the FastRCS algorithm 
$\eqref{mcs:fmcs}$.  
An important characteristic of FastRCS 
is that it can detect exact fit situations:
when $h$ or more observations lie exactly on
 a subspace, FastRCS will return the indexes 
of an $h$-subset of those observations and 
the hyperplane fitted by FastRCS will coincide
 with the subspace.
 
For each of the $M_p$ starting subset $H^m_0$, %%this also in fasthcs
Step $b$ grows the size of the corresponding $H^m_l$ from $p+1$ when 
$H^m_l=H^m_0$ to its final size ($h$) in $L$ steps, 
rather than in one as is done in FastLTS. We find that 
this improves the robustness of the algorithm when 
outliers are close to the good data. We also find that 
increasing $L$ does not improve performance much if $L$ 
is greater than 3 and use $L=3$ as default.

\vskip0.25cm
\hrule
\vskip0.1cm
\noindent \begin{equation}\label{mcs:fmcs}
\text{Algorithm FastRCS}
\end{equation}
\hrule
\begin{tabbing}
\=for \= $m=1$ to $M_p$ do:\\ 
$a$:\>\> 
\begin{math}
  H^m_0\gets\{\text{random } $(p+1)$-\text{subset}\}	
\end{math} \\
   $b$:\>\>for \=$l=1$ to $L$ do:\\
\>\>\>
\begin{math}
   R_i(H^m_l) \gets \displaystyle\ave_{k=1}^{K}  
    \frac{r_{i}^2(\tilde{\pmb\theta}^{mk})}
         {\displaystyle\ave_{j\in H^m_l}r_{j}^2(\tilde{\pmb\theta}^{mk})}
 \;\;\;1\leq i \leq n
\end{math}\\
\>\>\>set 
\begin{math}
  q \gets \lceil(n-p-1)l/(2L)\rceil+p+1			
\end{math}\\
\>\>\>set 
\begin{math}
  H^m_l \gets \left\{i: R_i(H^m_l) \leqslant R_{(q)}(H^m_l)\right\}
\end{math}\\
\>\>\>     \text{ (`growing step')} \\
\>\>end for\\
\>\>
\begin{math}
H^m \gets H^m_l
\end{math}\\
   $c$:\>\>  compute
\begin{math}
  I(H^m) \gets \displaystyle\ave_{k=1}^{K} I(H^m,\tilde{\pmb\theta}^{mk})
\end{math}
\\
\>end for\\
Keep $H^*$, the subset $H^m$ with lowest $I(H^m)$. 
\end{tabbing}
\vskip-0.1cm
\hrule
\vskip0.25cm

Empirically also, we found that small values 
for $K$, the number of elements of $\tilde{B}_K(H^m)$, 
is sufficient to achieve good results
 and that we do not gain much by increasing $K$
 above 25, so we set $K=25$ as the default. 
That such a small number of random regressions  
suffice to reliably identify the outliers is 
remarkable. This is because the hyperplanes 
used in FastRCS are fitted to $p$ 
observations drawn from the members of 
$H^m$ rather than, say, indiscriminately from among 
the entire set of data-points.   
Our choice always ensures a wider spread of 
directions when $H^m$ is uncontaminated and
 this yields better results.

Finally, in order to improve its small sample accuracy, 
we add a re-weighting step to our algorithm.
In essence, this re-weighting strives to award some
 weight to those observations lying close enough 
to the model fitted to the members of $H^*$. 
The motivation is that, typically, the 
re-weighted fit will encompass a greater share 
of the uncontaminated data.
 Over the years, many re-weighting 
procedures have been proposed \citep{mcs:MMY06}. 
The simplest is the so called
 one step re-weighting \citep[pg 202]{mcs:RL87}.  
Given an optimal $h$-subset $H^*$,  we get the final 
FastRCS parameters by fitting Model 
$\eqref{mcs:lm}$ to the members of  
\begin{eqnarray}\label{mcs:hp}
H^*_+=\{i:r_i(\pmb\theta^*)/(\Phi^{-1}(0.75)\med_{i=1}^n r_i(\pmb\theta^*))\leqslant2.5\}
\end{eqnarray}
FastLTS and FastS also use a re-weighting step (\cite{mcs:RV06} and \cite{rcs:Y87}) and, 
for all three algorithms, we will refer to the 
raw estimates as $(\pmb\theta^*,\sigma^2_*)$
and to the final, re-weighted vector of fitted
coefficients as $(\hat{\pmb\theta},\hat{\sigma}^2)$.

 Like FastLTS and FastS, FastRCS uses many
 random $(p+1)$-subsets as starting points. 
The number of initial $(p+1)$-subsets, $M_p$, 
must be large enough to ensure that at 
least one of them is uncontaminated. 
For FastLTS and FastS, for each starting 
$(p+1)$-subset, the computational complexity 
scales as $O(p^3+np^2)$ --comparable to 
 FastRCS for which it is $O(p^3+np)$.
The value of $M_p$ (and therefore the computational complexity
 of all three algorithms) grows exponentially with $p$. The 
actual run times will depend on implementation choices 
but in our experience are comparable for all three.
In practice this means that all three methods become
 impractical for values of $p$ much larger than 25. 
This is somewhat mitigated by the fact that they all  
 belong to the class of so called 
`embarrassingly parallel' algorithms, i.e. 
their time complexity scales as the inverse
 of the number of processors meaning that 
they are particularly well suited to benefit
 from modern computing environments. To 
enhance user experience, we implemented 
FastRCS in \cpp~ code wrapped in a portable 
\texttt{R} package \citep{mcs:R} 
distributed through \texttt{CRAN} (package \texttt{FastRCS}).

\section{Empirical Comparison: Simulation Study}\label{mcs:s3}

In this section we evaluate the behavior of 
FastRCS numerically and contrast its performance 
to that of FastLTS and FastS. 
For all three, we used their respective 
\texttt{R} implementation (package \texttt{robustbase} 
for the last two and \texttt{FastRCS} for FastRCS) 
with default settings except for the number of starting
 subsets which for all algorithms we set according to Equation 
$\eqref{mcs:Ns}$ and the maximum number of iterations
 for FastS which we increased to 1000. Each algorithm 
 returns a vector of estimated parameters $\hat{\pmb\theta}$ as 
 well as a an $h$-subset (denoted $H^+$) derived from it:
 \begin{eqnarray}
H^+=\{i:r_i(\hat{\pmb\theta})\leqslant r_{(h)}(\hat{\pmb\theta})\}
\end{eqnarray}
Our evaluation criteria are the bias of $\pmb\theta^*$ 
and the  rate of misclassification of the outliers.  

\subsection{Bias}

Given a central model 
$\mathcal{F}_u$ and
 an arbitrary distribution 
$\mathcal{F}_c$ (the index $c$ stands for contamination), consider 
the contamination model:
\begin{eqnarray}
\mathcal{F}_{\varepsilon}=(1-\varepsilon)\mathcal{F}_u(y_u|\pmb{x}_u)+\varepsilon\mathcal{F}_c(y_c|\pmb{x}_c),
\end{eqnarray}
where $\varepsilon$ is the rate of contamination of the sample. 
The bias measures the differences between the coefficients fitted to $\mathcal{F}_{\varepsilon}$
and those governing $\mathcal{F}_u(y_u|\pmb{x}_u)$ and is defined as the norm \citep{mcs:Ma89}:	%K%
\begin{equation}\label{sbias}
\mbox{bias}(\hat{\pmb\theta},\pmb\theta)=\sqrt{(\hat{\pmb\theta}-\pmb\theta)'\mathrm{Var}(\pmb\theta)^{-1}(\hat{\pmb\theta}-\pmb\theta)},
\end{equation}
for an affine and regression equivariant algorithm, w.l.o.g., 
we can set $\mathrm{Var}(\pmb\theta)=\mathrm{Diag}(p)$ 
and  $\pmb\theta=\pmb 0_p$ so that $\eqref{sbias}$ reduces to $||\hat{\pmb\theta}||$ 
and we will use the shorthand $\mbox{bias}(\hat{\pmb\theta})$
to refer to it. 

Evaluating the bias of an algorithm is an empirical matter. 
For a given sample, it will depend on
  the rate of contamination and the distance
 separating the outliers from the good part 
of the data. The bias will also depends
 on the spatial configuration of the outliers 
(the choice of $\mathcal{F}_c$). Fortunately, 
for affine and regression equivariant algorithms
 the worst configurations of outliers (those 
 causing the largest biases) are 
known and so we can focus on these cases. 

\subsection{Misclassification rate}

We can also compare the algorithms
 in terms of rate of contamination 
of their final $H^+$, the subset of 
$h$ observations with smallest 
values of $r_i(\hat{\pmb\theta})$. 
Denoting $I_c$ the index %K%
 set of the contaminated observations, 
the misclassification rate is:
\begin{eqnarray}
\mbox{Mis.Rate}(I_c,H^+)=\slfrac{\sum_{i\in H^+}\mathcal{I}(i\in I_c)}{\sum_{i=1}^n\mathcal{I}(i\in I_c)}.
\end{eqnarray}
This measure is always in $[0,1]$, thus
 yielding results that are easier to 
compare across configurations of outliers
 and rates of contamination. A value of 1 means that 
 $H^+$ contains all the outliers. 
The main difference with the bias 
criterion is that the misclassification
 rate does not account for how disruptive
 the outliers are. For example, when the outliers are 
close to the good part of the data, 
it is possible for $\mbox{Mis.Rate}(I_c,H^+)$
 to be large without a commensurate 
increase in $\mbox{bias}(\hat{\pmb\theta})$.

\subsection{Outlier configurations}
We generate many contaminated data-sets
 $(\pmb{X},\pmb{y})$ of size 
$n$ with $(\pmb{X},\pmb{y})=(\pmb{X}_{u},\pmb{y}_u)\cup (\pmb{X}_{c},\pmb{y}_c)$
 where $(\pmb{X}_{u},\pmb{y}_u)$ and $(\pmb{X}_{c},\pmb{y}_c)$ are, respectively, 
the genuine and outlying part of the sample. 
For equivariant algorithms, the worst-case configurations are known.
 These are the configurations of outliers that are, in a well defined sense, 
the most harmful. In increasing order of difficulty these are:

\begin{itemize}
\item Shift configuration. If we constrain the adversary to (a) set
        $\sigma^2\ge\mathrm{Var}(y_c|\pmb{x})$
	and (b) place the $y_c|\pmb{x}$ at a distance $\nu$ of 
	$\mathrm{E}(y_u|\pmb{x})$. Then, the adversary will set 
	$\mathrm{Var}(y_c|\pmb{x})=\sigma^2$ (Theorem 1 in 
       \citep{mcs:RW96}) and $\nu$ in order to satisfy (b).
      Intuitively, this makes the components of the mixture the
       least distinguishable from one another.
\item Point-mass configuration. If we omit constraint (a) 
      above but keep (b), the adversary will place $\pmb{X}_c$ on
      a subspace so that $\mathrm{Var}(y_c|\pmb{x})=0$ 
	(Theorem 2 in \citep{mcs:RW96}). Intuitively, this maximizes
	the cost of misidentifying any single outlier.
\end{itemize}

We can generate the $\pmb{X}_u$'s and the $\epsilon_i$'s
from standard normal distributions since all methods
 under consideration are affine and regression equivariant. 
Likewise, the $\pmb{X}_c$'s are draws from a multivariate normal distribution 
with $\mathrm{Var}(\pmb{X}_c)$ equal to $\mathrm{Diag}(p)$ (Shift) or $10^{-4}\mathrm{Diag}(p)$ (Point-mass) 
and $\mathrm{E}(\pmb{X}_c)$  set so that $\min_{i\in I_c}||\pmb x_i||=d_x\sqrt{\chi^2_{0.95;p-1}}$
 where $d_x$ is either one of 2 (nearby outliers) or 8 (far away outliers). 
Finally, $\mathrm{Var}(y_c|\pmb{x})$ is one of $\sigma^2$ or $10^{-4}\sigma^2$ 
depending again on whether the outlier configuration is Shift or Point-mass.
Next, for a given value of the parameters in Model 
$\eqref{mcs:lm}$ and a data matrix $\pmb{X}_c$, 
the bias will depend 
on the vertical distance between the outliers and 
the genuine observations. We will place 
the outliers such that they lie 
at a distance $\nu$ of the good data:
\begin{equation}\label{mcs:nu}
\nu=\min_{i\in I_c}\frac{|y_i-\mathrm{E}(y_{u}|\pmb{x}_i)|}{\mathrm{W}(\pmb{x}_i)}\,\,,
\end{equation}
where $\mathrm{W}(\pmb{x}_i)$ is 
half the asymptotic width of the usual LS prediction 
interval, evaluated at $\pmb{x}_i$. 
The complete list of simulation 
parameters follows:
\begin{itemize}
\item the dimension $p$ is one of $\{4,8,12,16\}$ and the sample size is $n=25p$, 
\item the configuration of the outliers is either
      Shift or Point-mass.
\item $\alpha\in\{0.5,0.75\}$ is the proportion of the sample 
that can be assumed to follow Model $\eqref{mcs:lm}$. In Section~\ref{mcs:sra} (Section~\ref{mcs:srb}) we consider the case where we set $\alpha=0.5$ ($\alpha=0.75$). 
\item $\varepsilon\in\{0.1,0.2,0.3,0.4\}$ (when $\alpha=0.5$) or $\varepsilon\in\{0.1,0.2\}$ (when $\alpha=0.75$).
\item the distance separating the outliers from the good data on the design 
space is $d_x=\{2,8\}$. The distance separating the outliers from $\pmb x'\pmb\theta$ is $\nu=\{1,2,...,10\}$. 
\item the number of initial $(p+1)$-subsets $M_p$ is given by \citep{mcs:MMY06}
      \begin{equation}\label{mcs:Ns}
        M_p=\log(0.01)/\log(1-(1-\varepsilon_0)^{p+1})\;,
      \end{equation}
      with $\varepsilon_0=4(1-\alpha)/5$ so that the probability of getting at least one
      uncontaminated starting point is always at least 99 percent.
\end{itemize}
We also considered other 
configurations such as so-called vertical outliers ($d_x=0$) 
or where $\nu$ is extremely large (i.e. $\ge1000$), 
but they posed little challenge for any of the algorithms, so we do 
not discuss these results.
In Figures~\ref{rcs:f1} to~\ref{rcs:f6} we display
 the bias (left panel) and the misclassification rate 
(right panel) for discrete combinations
 of the dimension $p$, contamination rate 
$\varepsilon$ and the degree of separation 
between the outliers and the genuine 
observations on the design space (which
 we control through the parameter $d_x$).
In all cases, we expect the outlier detection
 problem to become monotonically harder as
 we increase $p$ and $\varepsilon$.
Furthermore, undetected outliers that are located 
far away from the good data on the design space
will have more leverage on the fitted coefficients. 
For that reason, we also expect the biases 
to increase monotonically with $d_x$. 
Therefore, not much information will be lost by 
considering a discrete grid of a few 
values for these parameters.
 The configurations also depend on the distance separating
 the outliers from the true model, which we control 
through the simulation parameter $\nu$. 
 The effects of $\nu$ on the bias are harder to foresee: clearly 
  nearby outliers will be harder to detect but misclassifying 
distant outliers will increase the bias more. Therefore, we 
 will test the algorithms for many values (and chart 
the results as a function) of $\nu$. For both the bias and the
 misclassification curves, for each algorithm, a solid colored
 line will depict the median and a dotted line (of the same 
color) the 75th percentile. Here, each panel will be based on 1000 
simulations. 

\subsection{Simulation results (a)}\label{mcs:sra}

The first part of the simulation study covers 
the case where there is no information about the
 extent to which the data is contaminated. Then,
 for each algorithm, we have to set the size of the
active subset to $h$, corresponding to the lower bound 
of slightly more than half of the data. For FastLTS and 
FastRCS there is a single parameter $\alpha$ controling 
the size of the active subset so that we set $\alpha=0.5$. 
For FastS, we follow \citep[table 19, p. 142]{mcs:RL87} and set 
the value of the tunning parameters to $(b,c)=(0.5,1.547)$. 
\begin{figure*}[h!]
\centering
\includegraphics[width=0.49\textwidth]{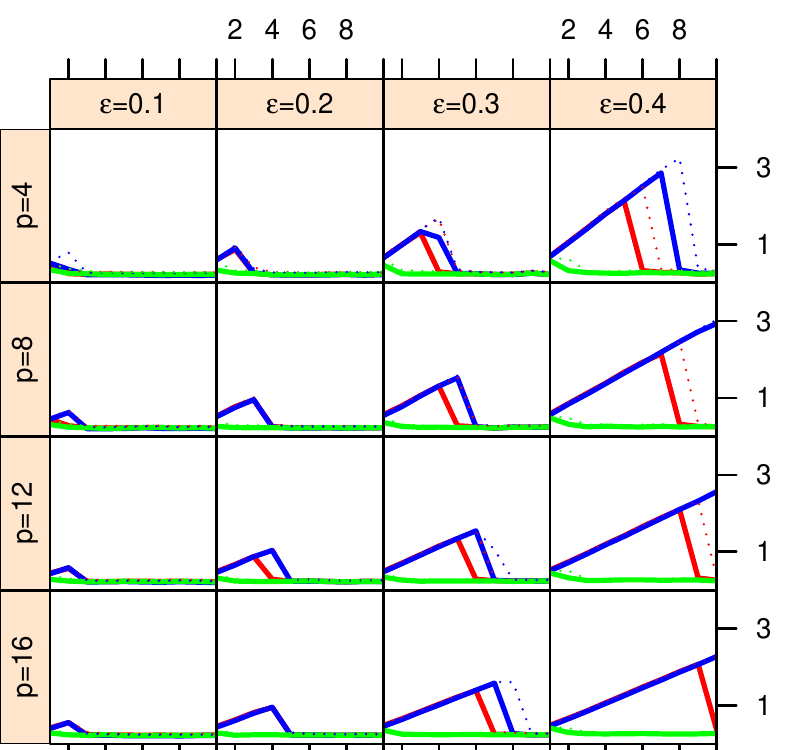}
\includegraphics[width=0.49\textwidth]{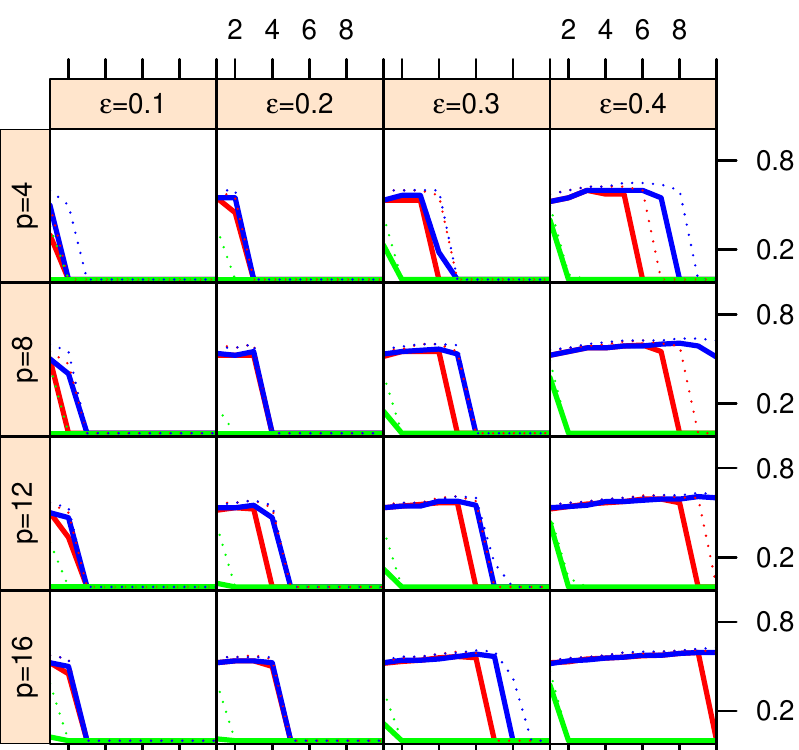}
\caption{$\mbox{Bias}(\hat{\pmb\theta})$ (left) and $\mbox{Mis.Rate}(I_c,H^+)$ (right)
	 for Shift contamination, 
         $\varepsilon=\{0.1,\ldots,0.4\}$, $p=\{4,\ldots,16\}$,  $d_x=2$, $\alpha=0.5$ 
	as a function of $\nu$.
         \textcolor{fmcd}{\underline{FastLTS}},
         \textcolor{fmve}{\underline{FastS}}, 
         \textcolor{RCS}{\underline{FastRCS}}.}
\label{rcs:f1}
\end{figure*}

In Figure~\ref{rcs:f1} we display the $\mbox{Mis.Rate}(I_c,H^+)$ and $\mbox{bias}(\hat{\pmb\theta})$   %K%
 curves of each algorithm at $d_x=2$ as a function of $\nu$ for different
values of $p$ and $\epsilon$ for the Shift configuration.
Starting at the second column, the $\mbox{Mis.Rate}(I_c,H^+)$ curves
are much higher for FastS and FastLTS than for FastRCS but the outliers included 
in their selected $h$-subsets do not exert enough pull on the fitted model to yield correspondingly 
large biases. From the the third column onwards, the outliers are now numerous 
enough to exert a visible pull on the coefficients fitted by FastLTS and FastS.  
By the fourth column of both panel, the performances of these two algorithms 
deteriorates further and they now fail to detect even outliers located very far from $\pmb x'\pmb\theta$. 
Far away outliers, if left unchecked, will exert a larger leverage on the $\hat{\pmb\theta}$
 and this is visible in the bias curves of FastLTS and FastS which are now much 
higher. The performance of FastRCS, on the other hand, is not affected by $\nu$,
 $p$ or $\varepsilon$ and remains comparable throughout. 
\begin{figure*}[h!]
\centering
\includegraphics[width=0.49\textwidth]{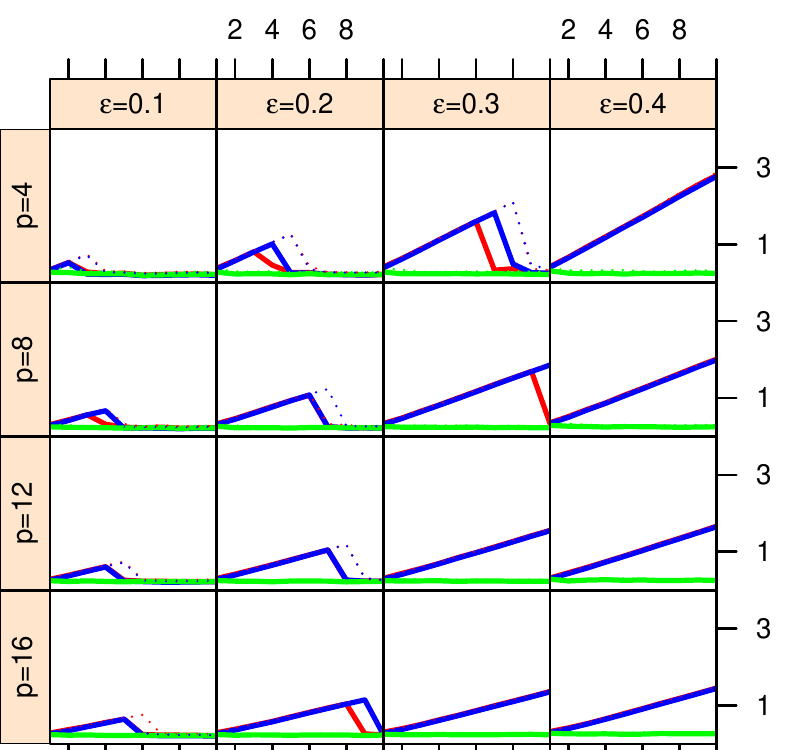}
\includegraphics[width=0.49\textwidth]{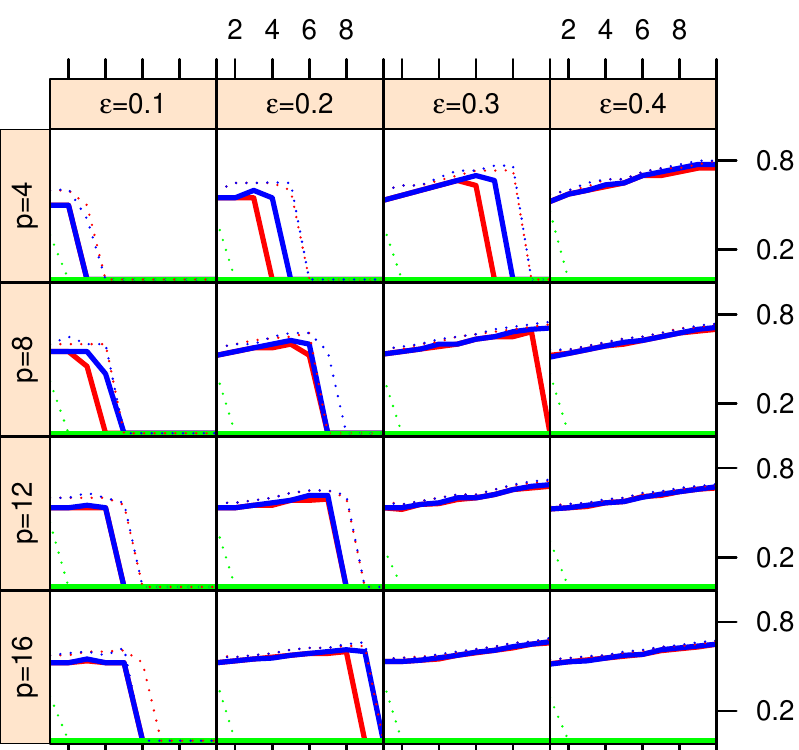}
\caption{$\mbox{Bias}(\hat{\pmb\theta})$ (left) and $\mbox{Mis.Rate}(I_c,H^+)$ (right) for
	 Shift contamination, $\varepsilon=\{0.1,\ldots,0.4\}$, $p=\{4,\ldots,16\}$, 
	 $d_x=8$, $\alpha=0.5$ as a function of $\nu$.
         \textcolor{fmcd}{\underline{FastLTS}},
         \textcolor{fmve}{\underline{FastS}}, 
         \textcolor{RCS}{\underline{FastRCS}}.}
\label{rcs:f2}
\end{figure*}

In Figure~\ref{rcs:f2}, we again examine the effects of Shift contamination, 
but for $d_x=8$. Now, starting at $\varepsilon=0.2$ and $p=12$ the  
$\mbox{Mis.Rate}(I_c,H^+)$ curves of FastLTS and FastS show that these
algorithms can not seclude even those outliers located within a rather 
large slab around $\pmb x'\pmb\theta$. Furthermore, from $\varepsilon\ge0.3$, 
as the configurations get harder, the maximum of the bias curves  
 clearly show that the parameters fitted by these two algorithms 
 diverge from $\pmb\theta$ in an increasingly large range of 
values of $\nu$.
Comparing the $\mbox{Mis.Rate}(I_c,H^+)$ curves  of FastLTS and FastS 
in Figure~\ref{rcs:f2} with those in Figure~\ref{rcs:f1},  we see 
that these two algorithms are noticeably less successful at identifying
 outliers far removed on the design space than nearer ones.
In contrast, in Figure~\ref{rcs:f2}, we see that the performance of
 FastRCS is very good both in terms of $\mbox{bias}(\hat{\pmb\theta})$
 and $\mbox{Mis.Rate}(I_c,H^+)$. Furthermore, these measures 
of performance remain consistently good across the different values 
of $p$, $\varepsilon$ and $\nu$. 
Comparing the performance of FastRCS in Figure~\ref{rcs:f2} with the results shown
 in Figure~\ref{rcs:f1} we also see that our algorithm is
 unaffected by the greater separation between the outliers 
and the good data in the $\pmb X$ space. 
\begin{figure*}[h!]
\centering
\includegraphics[width=0.49\textwidth]{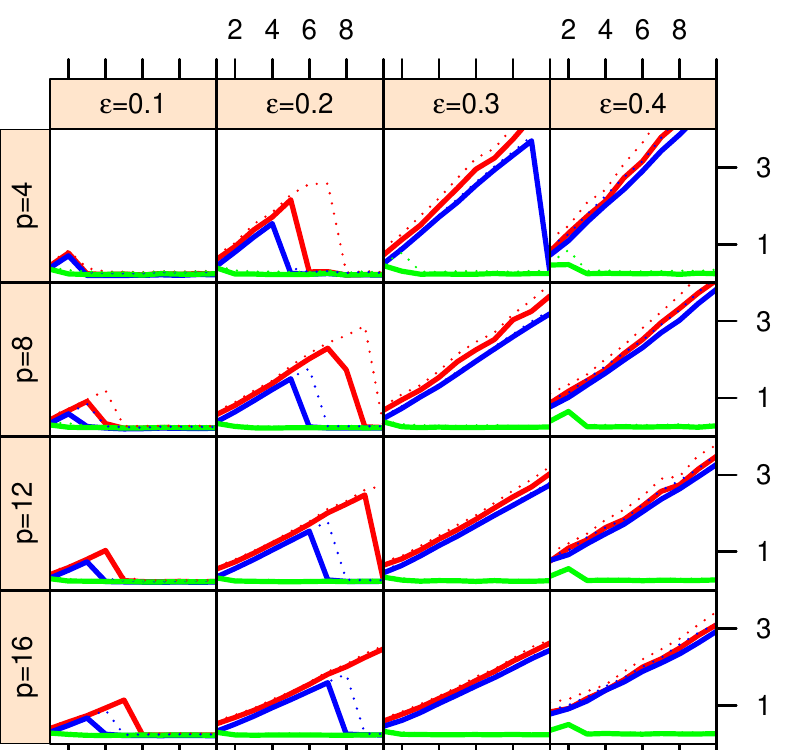}
\includegraphics[width=0.49\textwidth]{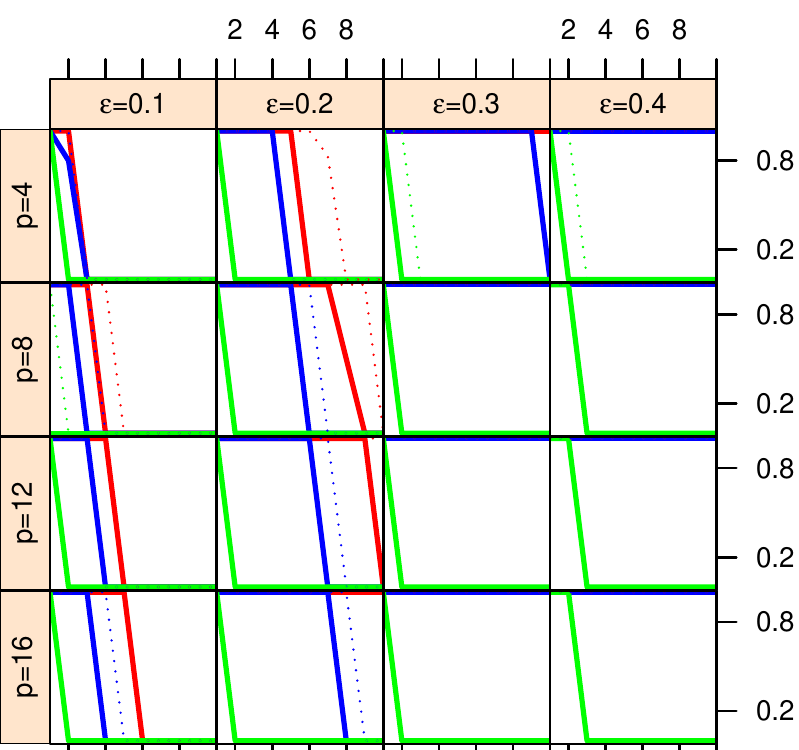}
\caption{$\mbox{Bias}(\hat{\pmb\theta})$ (left) and $\mbox{Mis.Rate}(I_c,H^+)$e (right) for Point-mass contamination, $\varepsilon=\{0.1,\ldots,0.4\}$, $p=\{4,\ldots,16\}$,  
	$d_x=2$, $\alpha=0.5$ as a function of $\nu$.
         \textcolor{fmcd}{\underline{FastLTS}},
         \textcolor{fmve}{\underline{FastS}}, 
         \textcolor{RCS}{\underline{FastRCS}}.}
\label{rcs:f3}
\end{figure*}

In Figure~\ref{rcs:f3}, we show the results for the more difficult case
 of Point-mass contamination with $d_x=2$. As expected, we find
 that concentrated outliers are causing much higher biases for 
FastS and FastLTS, especially in higher dimensions. Already when $\varepsilon=0.2$,
 FastS displays biases that are sensibly higher than their maximum values against 
the Shift configuration. From $p=8$, both algorithms also yields contaminated 
$h$-subsets for most values of $\nu$.  
Looking at the $\mbox{Mis.Rate}(I_c,H^+)$ curves of these two algorithm, we 
also see that in many of cases, the optimal $h$-subsets selected by FastLTS 
and FastS actually contain a higher fraction of outliers than the original sample. 
In contrast, for FastRCS, the performance curves in  
Figure~\ref{rcs:f3} are essentially similar to those shown in 
  Figure~\ref{rcs:f1}, attesting again that our algorithm is not unduly 
affected by the spatial concentration of the outliers. 
\begin{figure*}[h!]
\centering
\includegraphics[width=0.49\textwidth]{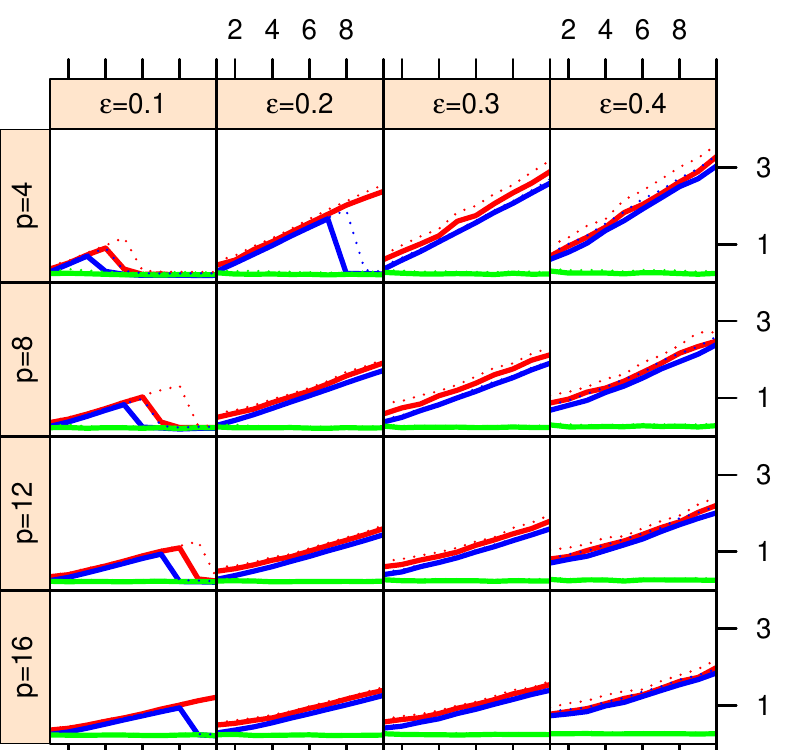}
\includegraphics[width=0.49\textwidth]{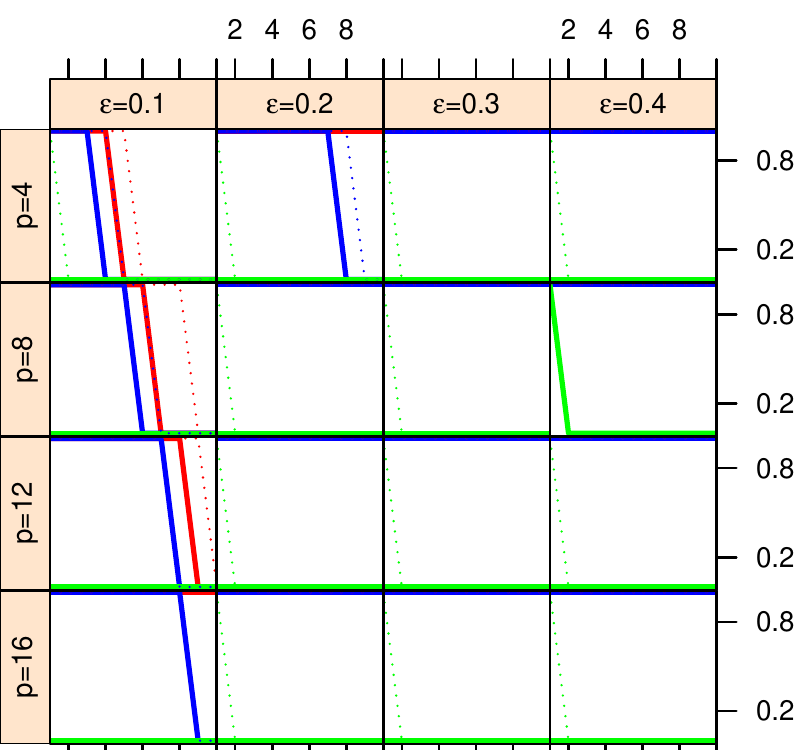}
\caption{$\mbox{Bias}(\hat{\pmb\theta})$ (left) and $\mbox{Mis.Rate}(I_c,H^+)$ (right) for Point-mass contamination,  $\varepsilon=\{0.1,\ldots,0.4\}$, $p=\{4,\ldots,16\}$,  
	 $d_x=8$, $\alpha=0.5$  as a function of $\nu$.
         \textcolor{fmcd}{\underline{FastLTS}},
         \textcolor{fmve}{\underline{FastS}}, 
         \textcolor{RCS}{\underline{FastRCS}}.}
\label{rcs:f4}
\end{figure*}

In Figure~\ref{rcs:f4}, we examine the effects of	
 Point-mass contamination for $d_x=8$. As in the case
 for Shift contamination, we see that an increase in $d_x$
 entails a decline in performance for both FastLTS and FastS with
 high maximum values of the bias curves now starting at 
$\varepsilon\ge0.2$ already. In terms of $\mbox{Mis.Rate}(I_c,H^+)$ curves, 
too, both algorithm have their weakest showings, with selected subsets 
that have a higher rate of contamination that the original data-sets for most 
values of $\nu$ starting from the middle rows of the first column already.
As before, the lower rate of outlier detection combined with 
the greater separation of the outliers on the design space means that the 
bias curves of FastLTS and FastS are worse than those shown 
in Figure~\ref{rcs:f3}.
Again, contrast this with the behavior of FastRCS which maintains 
 low and constant bias and misclassification curves throughout. 
Further comparing the performance curves of FastRCS for the configuration 
considered in Figure~\ref{rcs:f4} to those considered in Figures~\ref{rcs:f3} 
and~\ref{rcs:f2} we see, again, that FastRCS is not affected by 
either the spatial concentration or the degree of separation of 
the outliers.  

\subsection{Simulation results (b)}\label{mcs:srb}

In this section, we now consider the case, important 
in practice, where the user can confidently place 
an upper bound on the rate of contamination of the 
sample. To fix ideas, we will set
  the proportion of the sample assumed
 to follow Model $\eqref{mcs:lm}$, to approximately $3/4$ so that in this 
section $\#H^*\approx 3h/2$.
 For FastLTS  and FastRCS we adapt the algorithms by setting their 
respective $\alpha$ parameter to 0.75. For FastS, 
we again follow \citep[table 19, p. 142]{mcs:RL87} and now set $(b,c)=(0.75,2.937)$.
For all three algorithms we also reduce the number of
 starting subsets by setting $\varepsilon_0=0.2$ in Equation $\eqref{mcs:Ns}$.
 Then, as before, we will measure the effects of the various configurations of
 outliers on the algorithms (but now for $\varepsilon\in\{0.1,0.2\}$).

\begin{figure*}[h!]
\centering
\includegraphics[width=0.49\textwidth]{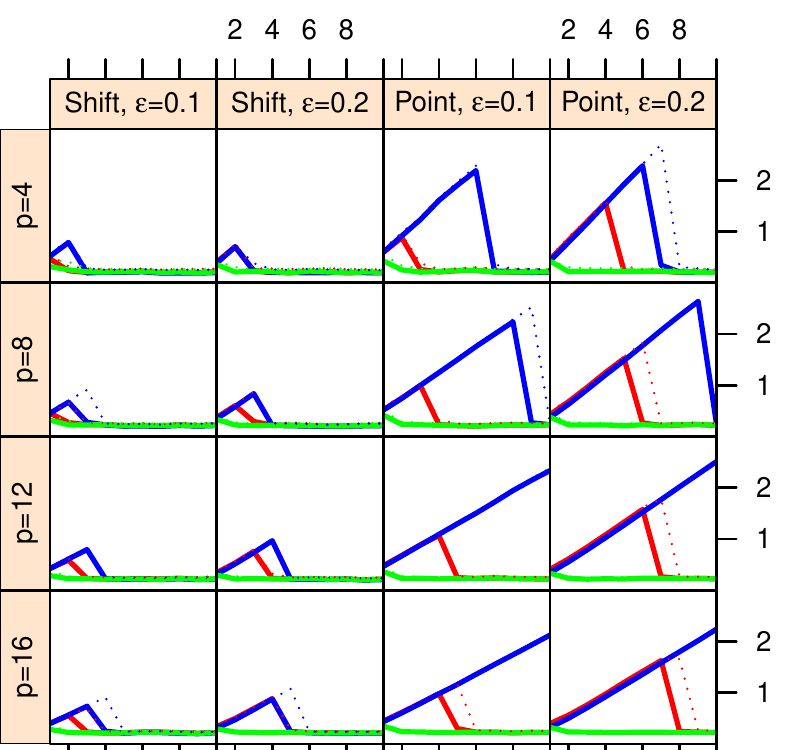}
\includegraphics[width=0.49\textwidth]{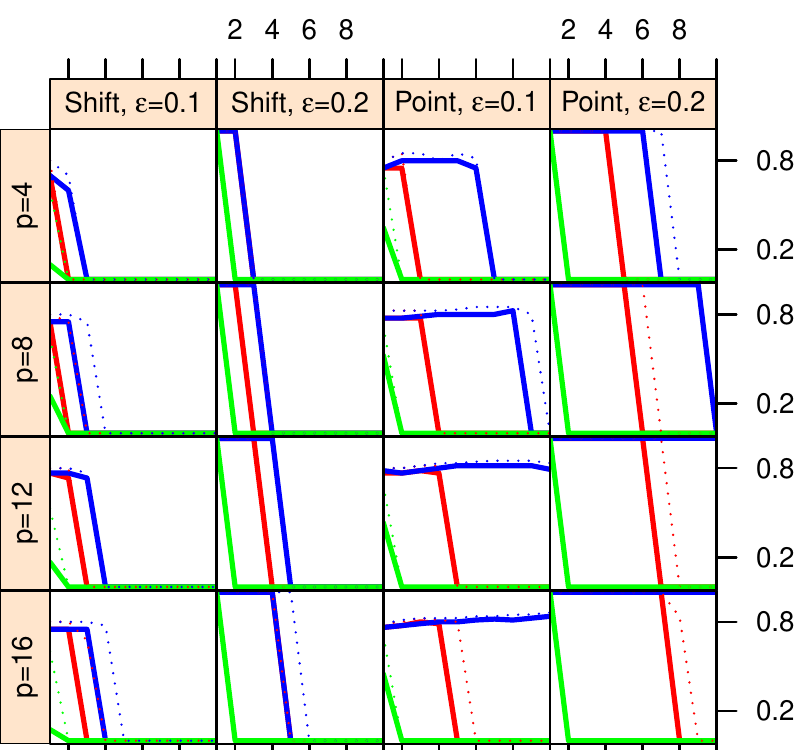}
\caption{$\mbox{Bias}(\hat{\pmb\theta})$ (left) and $\mbox{Mis.Rate}(I_c,H^+)$ (right) for Shift and Point-mass contamination, $\varepsilon=\{0.1,0.2\}$, $p=\{4,\ldots,16\}$, $\alpha=0.75$, $d_x=2$ as a function of $\nu$.
         \textcolor{fmcd}{\underline{FastLTS}},
         \textcolor{fmve}{\underline{FastS}}, 
         \textcolor{RCS}{\underline{FastRCS}}.}
\label{rcs:f5}
\end{figure*}

Figure~\ref{rcs:f5} summarizes the case where $d_x=2$. 
The first two columns of $\mbox{bias}(\hat{\pmb\theta})$ (left) 
and $\mbox{Mis.Rate}(I_c,H^+)$ contain results for the Shift 
configuration, and the last two, for Point-mass. 
For the Shift configuration, when the outliers are far from
the hyperplane fitting the good 
part of the data, all methods have low biases. Though FastS and 
FastLTS still deliver optimal $h$-subsets that are heavily 
contaminated by outliers, these do not exert a large enough 
pull on the $\hat{\pmb\theta}$ to affect the biases.
Generally all methods yield results that are comparable with the 
corresponding cases shown in Figure~\ref{rcs:f1}. In the case of Point-mass 
contamination (shown in the last two columns), FastLTS and FastS
again experience greater difficulty with the Point-mass 
configuration and results are worse than those shown in 
Figure~\ref{rcs:f3} for the corresponding settings. 
Again, the results for FastRCS remains similar
 to those depicted in Figure~\ref{rcs:f3}.
\begin{figure*}[h!]
\centering
\includegraphics[width=0.49\textwidth]{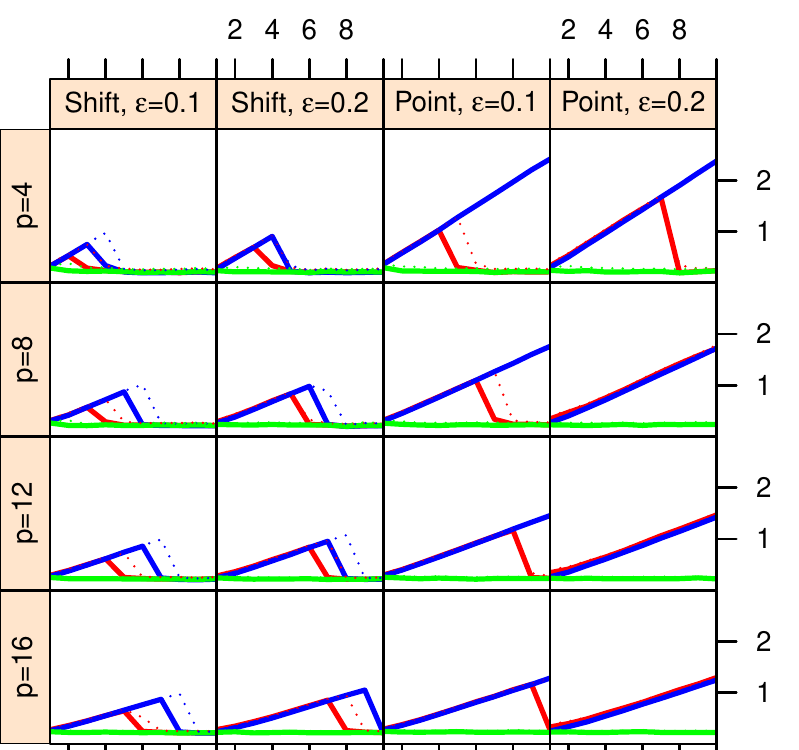}
\includegraphics[width=0.49\textwidth]{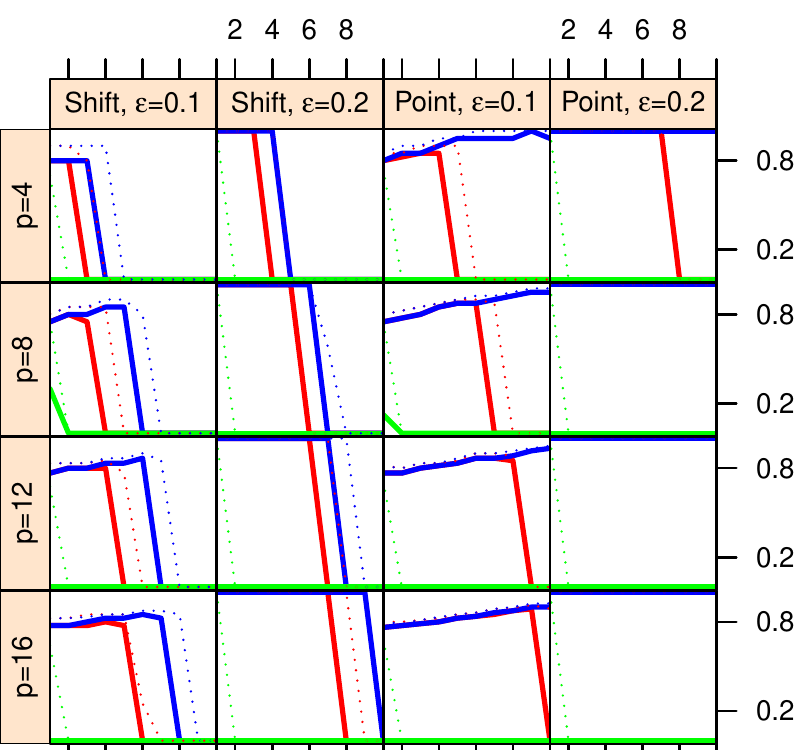}
\caption{$\mbox{Bias}(\hat{\pmb\theta})$ (left) and $\mbox{Mis.Rate}(I_c,H^+)$ (right) for Shift and Point-mass contamination,
         $\varepsilon=\{0.1,0.2\}$, $p=\{4,\ldots,16\}$, $\alpha=0.75$, $d_x=8$ 
	 as a function of $\nu$.
         \textcolor{fmcd}{\underline{FastLTS}},
         \textcolor{fmve}{\underline{FastS}}, 
         \textcolor{RCS}{\underline{FastRCS}}.}
\label{rcs:f6}
\end{figure*}

Figure~\ref{rcs:f6} depicts the simulation results for $\alpha=0.75$ when
 $d_x=8$. As before, this places additional strain on FastLTS and FastS.
 In the case of Shift contamination, the results for these two algorithms 
are qualitatively the same as in  Figure~\ref{rcs:f2}. For FastLTS and FastS, 
the results are worst for Point-mass contamination and show, again, a clear 
deterioration vis-\'a-vis Figure~\ref{rcs:f4}.  Now, 
FastLTS fails to identify the outliers in any but the easiest configuration, 
and FastS performs poorly in all configurations.
In contrast, FastRCS yields the same performance, whether
 it is measured in terms of $\mbox{bias}(\hat{\pmb\theta})$ or
 $\mbox{Mis.Rate}(I_c,H^+)$, as those obtained in the 
same configurations, but with a higher value of $\alpha$. 

Overall in our tests, we observed that FastLTS and FastS both
exhibit considerable strain. In many situations these two 
algorithms yield optimal $h$-subsets that are so heavily contaminated 
 as to render them wholly unreliable for finding the outliers.
 In these situations, both algorithm also invariably yield
 vectors of fitted parameters $\hat{\pmb\theta}$ that are completely 
swayed by the outliers and, consequently, very poorly fit the 
genuine observations.
We note that as the configurations become more challenging, 
 the bias curves for FastLTS and FastS deviate upward, and 
the $\mbox{Mis.Rate}(I_c,H^+)$ curves advance to the right 
in a systemic manner. We observe no corresponding effect for
 FastRCS. These quantitative differences between FastRCS and 
the other two algorithms, repeated over adversary 
configurations, lead us to interpret them as indicative of a 
qualitative difference in  robustness.

\section{Empirical Comparison: Case Studies}\label{mcs:s4}

In the previous section, we compared FastRCS to two state 
of the art outlier detection algorithms in situations that
 were designed to be most challenging for equivariant procedures.  
In this section, we will also compare FastRCS to FastLTS 
and FastS, but this time using two real data examples. 
The first illustrate the use of FastRCS in 
situations where $n/p$ is small (typically these situations
 are particularly challenging for anomaly detection procedures) 
while the second example illustrates the use of FastRCS in 
situations where $p$ is large.

\subsection{Slump Data}

In this section, we
consider a real data problem from the
 field of engineering: the Concrete "Slump"
Test Data set \citep{mcs:Y07}. 
This data-set consists of 7 input variables 
measuring the quantity of cement, fly ash,
 blast furnace slag,
water, super-plasticizer, coarse aggregate,
 and fine aggregate used
 to make the corresponding variety
 of concrete. Finally, we use 28-day
 compressive strength as
 response variable.
The "Slump" data-set is actually composed of two class 
 of observations collected over two separate
periods with the first set of measurements predating the 
second ones by several years. After excluding several 
data-points that are anomalous in that the values
 of either the slag and fly ash variable is
exactly zero, we are left with 59 measurements, with 
first 35 data (last 24) points belonging to the set $J_O$ ($J_N$)
of older (newer) observations. In this exercise, we will hide
the vector of class labels, in effect casting the members of 
$J_N$ as the outliers and the task of all the algorithms will 
be to reveal them.

To ensure reproducibility we ran all algorithms 
 with option \texttt{seed=1} and default values of
  the parameters (except for the number of starting subset 
  $M_p$ which we set according to Equation~\eqref{mcs:Ns}) and
  included both data-sets used in this section in the FastRCS package. 
For all three algorithms, we will denote as 
 $\hat{\pmb\theta}_j$ the fit found by algorithm $j$, 
 $\hat{H}_{j}=\{i:r_i(\hat{\pmb\theta}_j)/\hat{\sigma}_j\leqslant2.5\}$
  will denote the subset of observations classified as "good"  and 
$\hat{H}_{\setminus j}=\{\{1,\ldots,n\}\setminus\hat{H}_j\}$ its complement.
\begin{figure*}[h!]
\centering
\includegraphics[width=1\textwidth]{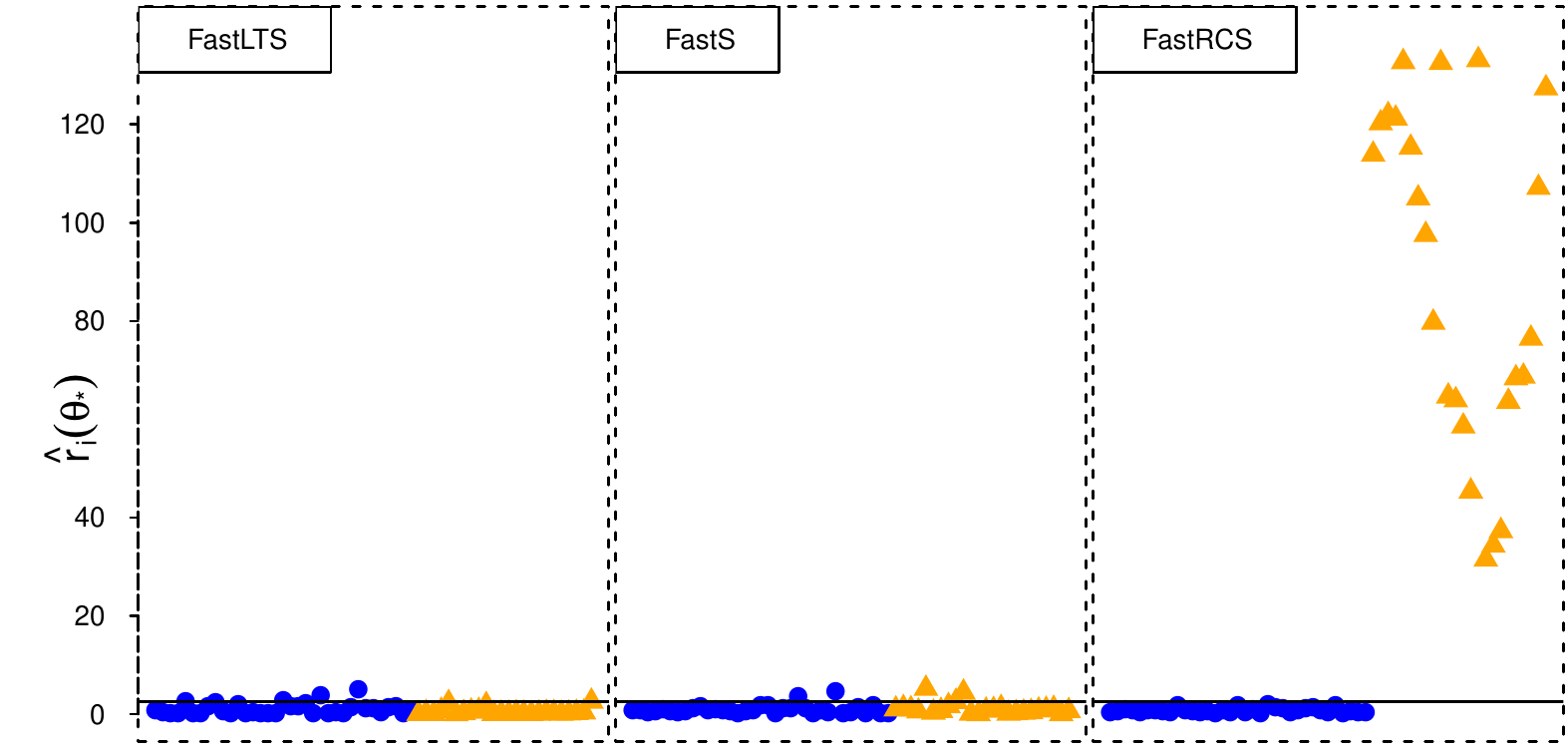}
\caption{Residual distances $r_i(\hat{\pmb\theta}_j)/\hat{\sigma}_j$ for the 
FastLTS (left), FastS (middle), and FastRCS (right), ran on
the "Slump" data-set. In each panel, 
the dark blue (light orange) dots (triangles) depict
the $r_i(\hat{\pmb\theta}_j)/\hat{\sigma}_j$ values for the members of $J_O$ ($J_N$).}
\label{mcs:f7}
\end{figure*}

Next, we ran all three algorithms on the "Slump" data-set, using $M_p=500$. 
In Figure \ref{mcs:f7}, we display the standardized residual distance 
$r_i(\hat{\pmb\theta}_j)/\hat{\sigma}_j$, for each algorithm in a separate panel.
The dark blue residual (light orange) points (triangles) depict the 
$r_i(\hat{\pmb\theta}_j)/\hat{\sigma}_j$ for the 35 members of $J_O$ (24 members of $J_N$).
An horizontal line at $2.5$ shows the usual outlier rejection threshold.
For both FastLTS and FastS, the values of the $r_i(\hat{\pmb\theta}_j)/\hat{\sigma}_j$ 
(shown in the left and middle panel of 
Figure~\ref{mcs:f7} respectively)
corresponding to the members of $J_O$ and $J_N$ largely overlap. As a result, in both cases, the optimal $h$-subset includes both members of $J_N$ and $J_O$. 
 Looking at the composition of the $H^+$ subset found by FastLTS, we see that it
 contains 19 (out of $h=34$) members of $J_N$. For FastS, $H^+$
 contains 13 members of $J_N$. 							%K%:removed §
Considering now the observations classified as good (the members of $\hat{H}_j$)	%K%
by both algorithms, we see that FastLTS (FastS) 
finds only 15 (5) outliers. The outliers found by both approaches are 
well separated from the fitted model (the nearest outlier lies at a
 standardized residuals distance of $2.9$ for both algorithms) but 
they tend to be located near the 	%K%
good observations on the design space. For example, the nearest 
 outlier lie at a Mahalanobis distance of $0.7\sqrt{\chi^2_{0.95,7}}$ 
wrt to the members of $\hat{H}_{\text{FastLTS}}$ and $0.6\sqrt{\chi^2_{0.95,7}}$
 wrt to the members of $\hat{H}_{\text{FastS}}$. Finally,
the members of the $\hat{H}_j$ subsets identified by 	%K%
both algorithms contains, again, many members of $J_N$: 21 (out of $|\hat{H}_{\text{FastLTS}}|=44$)
 for $\hat{H}_{\text{FastLTS}}$ and 21 (out of 54) for $\hat{H}_{\text{FastS}}$.	%K%
   
In contrast to the innocent looking residuals seen in the 
first two panels of Figure~\ref{mcs:f7}, the plot of the 
$r_i(\hat{\pmb\theta}_j)/\hat{\sigma}_j$ corresponding to the FastRCS fit (shown in rightmost panel), 
clearly reveals the presence of many observations that do not follow the multivariate pattern of the bulk of the data. 	%K% 
For example, the outliers identified by FastRCS are much more distinctly deviating from the pattern 
set by the majority of the data than those identified by FastS and FastLTS: the nearest outlier now
 stands at $r_i(\hat{\pmb\theta}_j)/\hat{\sigma}_j=32$. 
Furthermore, the composition of these two groups is now consistent 
with the history of the "Slump" data-set as the $H^+$ found by FastRCS is solely populated by members of $J_O$ and 
 $\hat{H}_{\text{FastRCS}}$ contains all the (and 
only those) 35 observations from $J_O$.
Interestingly, the two groups of observations identified by FastRCS are
 clearly distinct on the design space where they form two well separated 
cluster of roughly equal volume on the design space. 
For example, the closest member of $\hat{H}_{\setminus\text{FastRCS}}$ 
lies at a Mahalanobis distance of over $19\sqrt{\chi^2_{0.95,7}}$ 
wrt the members of $\hat{H}_{\text{FastRCS}}$.

The blending of members of both $J_N$ and $J_O$ in
both $\hat{H}_{\text{FastLTS}}$ and $\hat{H}_{\text{FastS}}$ 
despite the large separation between the members $J_N$ and those from $J_O$,
on the design space as well as to the hyperplane fitting
 the observations with indexes in $J_O$ 
 together suggest that the optimal subset
 selected by both FastLTS and FastS do in fact harbor many 
positively harmful outliers. 

\begin{comment}
\begin{table*}
  \centering
  \caption{: $t$-statistics associated with $\hat{\pmb\theta}_j$, "Slump" data-set.}
    \begin{tabular}{|r|r|r|r|r|r|r|r|r|}
    \hline
          & $\hat{\alpha}_j$    & $\hat{\beta}_{j1}$     & $\hat{\beta}_{j2}$     & $\hat{\beta}_{j3}$     &$\hat{\beta}_{j4}$     & $\hat{\beta}_{j5}$     & $\hat{\beta}_{j6}$     & $\hat{\beta}_{j7}$ \\
    \hline
\textbf{FastLTS}  & 2.8   & 0.4   & -2.3  & -0.1   & -4.2  & 0.2   & -2.9  & -2.1 \\
    \textbf{FastS}    & 4.3   & -2.0     & -3.7  & -2.3   & -5.0    & -2.0   & -4.2  & -4.0 \\
    \textbf{FastRCS}  & -2.9  & 3.3   & -0.1  & 2.5   & 2.7   & 3.1   & 2.9   & 2.9 \\
    \hline
    \end{tabular}%
\label{mcs:tab1}%
\end{table*}%
\end{comment}

Consecutively, because the outliers in this example 
 are well separated from the bulk of the data, the presence of even a 
handful of them in $\hat{H}_{\text{FastLTS}}$ ($\hat{H}_{\text{FastS}}$) 
indicates that they have pulled the resulting fits 
 so much in their direction as to render parameters fitted  
by both algorithms altogether unreliable. 

\subsection{The "Don't Get Kicked!" data-set}

\begin{figure*}[h!]
\centering
\includegraphics[width=1\textwidth]{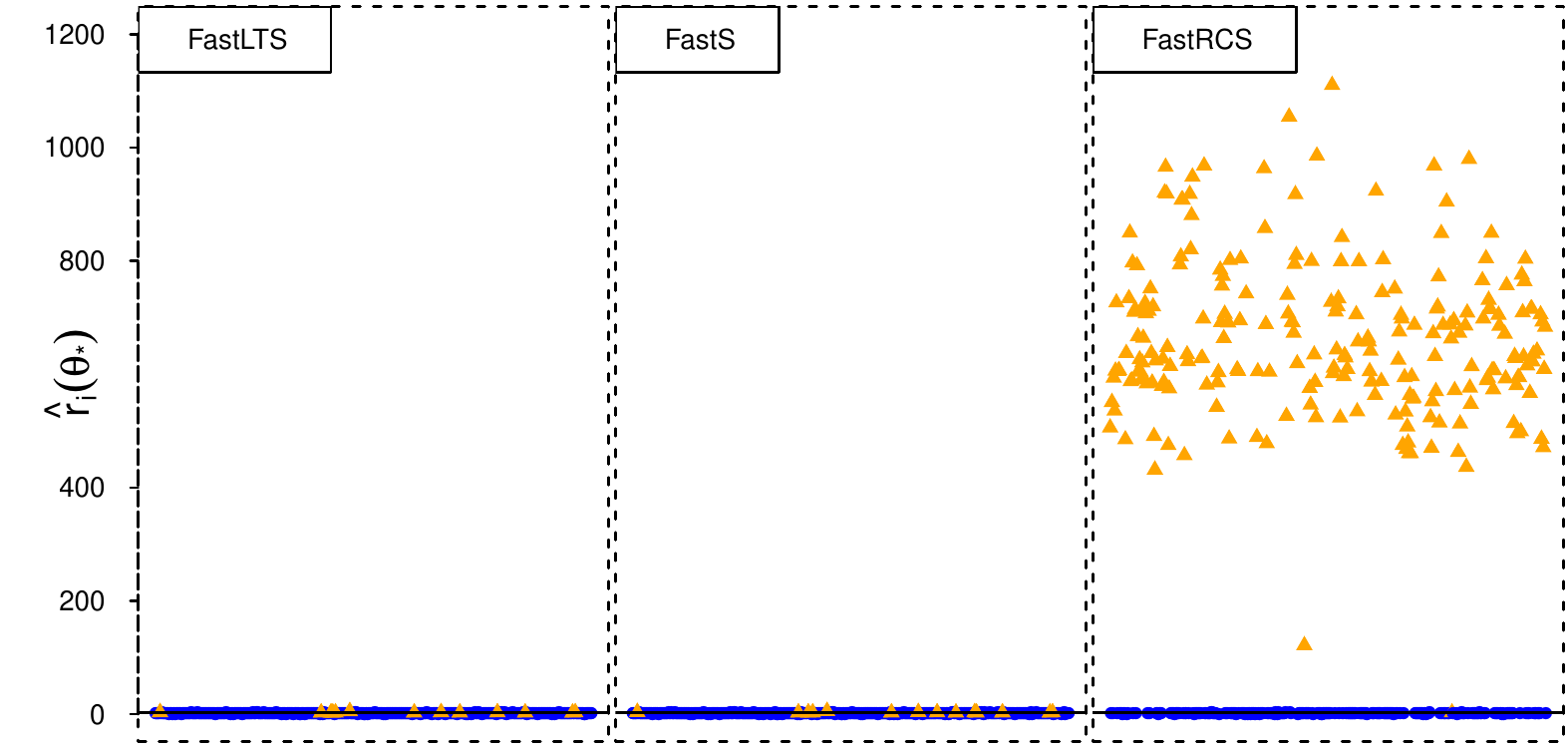}
\caption{Residual distances $r_i(\hat{\pmb\theta}_j)/\hat{\sigma}_j$ for the 
FastLTS (left), FastS (middle), and FastRCS (right), obtained 
on the "Don"t get kicked!" data-set. In each panel, 
the light orange triangles depict those observations for which
$r_i(\hat{\pmb\theta}_j)/\hat{\sigma}_j>2.5$.}
\label{mcs:f8}
\end{figure*}

In this subsection, we illustrate the behavior
of FastRCS on a second real data problem from
a commercial application: the "Don't Get Kicked!"
data-set~\citep{mcs:Ka12}. This data-set contains
thirty-four variables pertaining to the
resale value of 72,983 cars at auctions. Of these,
 we use of
the continuous ones to model the sale price of a car
 as a linear function of eight measures of current 
market acquisition prices, the number of miles on the
 odometer and the cost of the warranty. 
More specifically, we consider 
as an illustrative example all 488 data-points corresponding to sales of 
the Chrysler Town \& Country but the pattern we find below 
repeats for many other cars in this data-set.
We ran all three algorithms on the data-set (with $M_p=3000$) and, 
in Figure \ref{mcs:f8}, we plot the vector of fitted standardized 
residual distance $r_i(\hat{\pmb\theta}_j)/\hat{\sigma}_j$  returned 
by each in a separate panel together with, again, an horizontal line at $y=2.5$ demarcating 
the usual outlier rejection threshold. Finally, for each algorithm, we also show  
the observations flagged as influential as light orange triangles. 

We first discuss the results of FastLTS and FastS 
jointly (shown in the left and middle panel of 
Figure~\ref{mcs:f8} respectively) since they are broadly similar. 
Here again, the first two panels depict a reasonably well behaved 
data-set with very few harmful outliers.
Overall the number of data points for which
 $r_i(\hat{\pmb\theta}_j)/\hat{\sigma}_j$ is larger than 2.5 
is 13 for FastLTS and 14 for FastS. These are quiet close to 
the the fitted $\hat{\pmb\theta}$ and the smallest 
value of $r_i(\hat{\pmb\theta}_j)/\hat{\sigma}_j$ for 
the  outliers is $2.5$ for both algorithms 
(the second nearest lies at a standardized distance 
of $2.6$).
Furthermore, these outliers are close, on the design 	%K%
space, to the majority of the data.
For example, for both algorithms, the nearest outlier
 lies at a Mahalanobis distance of $0.4\sqrt{\chi^2_{0.95,10}}$ 
wrt to the good observations (the second closest is located at 
a distance of $0.5\sqrt{\chi^2_{0.95,10}}$). Overall, both algorithms 
suggest that observations forming the \textit{Don't Get Kicked!} 
data-set is broadly consistent with Model $\eqref{mcs:lm}$ except for 
$\approx2$\% of outliers. 

In this example again, we find that the FastRCS fit reveals a much more 
elaborate structure to this data-set. In this case, $\hat{H}_{\text{FastRCS}}$ is composed
 of 262 observations, barely more than $h=250$ which %erictype
imply a data-set beset by outliers.  Furthermore, even 
a cursory inspection of the standardized residuals clearly
 reveals the presence of a large group of observations not following 
 the multivariate pattern of the bulk of the data. 
Surprisingly in the light of the fit found by the other algorithms, 
we find that the outliers identified by FastRCS in this data-set are
actually very far from the model fitting the bulk of the data. 
Setting aside row 171 (visible on the right panel of Figure~\ref{mcs:f8} 
as the isolated outlier lying 
somewhat closer to the genuine observations), we find that the nearest
outlier lies at a standardized residual distance of over 120 wrt to the
FastRCS fit on the model space. Considering now the design space alone, we 
 find that here too the members of $\hat{H}_{\text{FastRCS}}$ are
 clearly separated from the bulk of the data: we find that the nearest 
 outlier (again setting observation 171 aside) lies at a Mahalanobis distance 
 of over $685\sqrt{\chi^2_{0.95,10}}$ wrt to the members of $\hat{H}_{\text{FastRCS}}$.

A salient feature of this exercise is that the outliers in the "Don't get kicked!" data-set
are genuine discoveries in the sense that when examining the variables
(including the qualitative ones not used in this analysis), we
 could not find any single one exposing them, yet they materially affect the model.
As with the "Slump" data-set, the "Don't Get Kicked!" data-set
is also interesting because the outliers identified
 (exclusively) by FastRCS are not  analogous to the worst 
case configurations we considered in the simulations. 
Indeed, far from resembling the concentrated outliers we
know to be most challenging for FastS and FastLTS,  
the members of $\hat{H}_{\setminus\text{
FastRCS}}$ seem to
form a scattered cloud of points, occupying a much
larger volume on the design space than the members
 of the main group. Nevertheless, here too, the 
 fit found by both FastLTS and FastS 
lumps together observations stemming from 
very disparate groups. In this case too, the large separation between the
members of $\hat{H}_{\text{FastRCS}}$ and those of
$\hat{H}_{\setminus\text{FastRCS}}$ along the
design space as well as wrt $\hat{\pmb\theta}_{\text{FastRCS}}$ 
and the fact that many observations flagged as outliers
are awarded weight in the fit found by both 
FastS and FastLTS together suggest that these
 data-points exert a substantial influence on the 
model fitted by these algorithms.
 Consequently, we do not
expect the coefficients fitted by either to accurately
describe, in the sense of Model $\eqref{mcs:lm}$,
any subset of the data. 

\begin{comment}
\begin{table*}
\centering
\caption{:$t$-statistics associated with $\hat{\pmb\theta}_j$, "Don't Get Kicked!"  data-set.}
    \begin{tabular}{|r|r|r|r|r|r|r|r|r|r|r|r|}
    \hline
             & $\hat{\alpha}_j$    & $\hat{\beta}_{j1}$     & $\hat{\beta}_{j2}$     & $\hat{\beta}_{j3}$     &$\hat{\beta}_{j4}$     & $\hat{\beta}_{j5}$     & $\hat{\beta}_{j6}$     & $\hat{\beta}_{j7}$    &$\hat{\beta}_{j8}$     & $\hat{\beta}_{j9}$     & $\hat{\beta}_{j10}$\\
    \hline
    \textbf{FastLTS}  & 20.8  & 3.0   & 1.3   & 1.2   & -2.0  & -1.6  & 0.1     & 0.0  & 0.5   & 1.7   & -7.0 \\
    \textbf{FastS}    & 20.6  & 2.7   & 1.4   & 1.1   & -1.7  & -1.3  & 0.1  & -0.1  & 0.2   & 1.8   & -7.0 \\
    \textbf{FastRCS}  & -0.8  & -0.4  & 0.8   & 0.9   & -0.8  & -2.3  & -0.1     & 0.4   & 1.5   & 1.7   & -4.9 \\
    \hline
    \end{tabular}%
\label{mcs:tab2}%
\end{table*}%
\end{comment}
% Table generated by Excel2LaTeX from sheet 'Sheet2'

\section{Outlook}
   
In this article we introduced RCS, 
a new outlyingness index and FastRCS,
 a fast and equivariant algorithm 
for computing it. Like many other outlier
 detection algorithms, the performance   
of FastRCS hinges crucially on correctly
 identifying an $h$-subset of uncontaminated
 observations. Our main contribution is to
 characterize this $h$-subset using a new
  measure of homogeneity based on residuals   
 obtained over many random regressions. 
This new characterization was designed to be insensitive to the 
configuration of the outliers.

Through simulations, we considered 
configurations of outliers that are
 worst-case for affine and regression 
equivariant algorithms, and found that
 FastRCS behaves notably better than the
 other procedures we considered, often
 revealing outliers that would not have
 been identified by the other approaches.   
In most applications, admittedly, contamination
 patterns will not always be as difficult as
 those we considered in our simulations and 
in many cases the different methods will, hopefully, concur. 
Nevertheless, using two real data examples we were able to 
establish that it is possible for real world situations  
to be sufficiently challenging as to push current state
 of the art outlier detection procedures to their  
limits and beyond, justifying the  
development of better solutions. %erictype 
In any case, given that in practice we do not
 know the configuration of the outliers, as
 data analysts, we prefer to carry our inferences
 while planing for the worst contingencies.

\singlespacing
\bibliographystyle{model2-names}

\end{document}